\documentclass[reprint,flushbottom]{revtex4-1}

\usepackage{color}

\usepackage[T1]{fontenc}
\usepackage[latin1]{inputenc}
\usepackage{bm}
\usepackage{amsmath}
\usepackage{amssymb}
\usepackage{mathrsfs}
\usepackage{latexsym}
\usepackage{amsfonts}
\usepackage{epsfig}
\usepackage{psfrag}

\newcommand{\f}[1]{\mbox{\boldmath$#1$}}

\newcommand{\na}{\mbox{\boldmath$\nabla$}}
\newcommand{\bea}{\begin{eqnarray}}
\newcommand{\ea}{\end{eqnarray}}
\newcommand{\bmult}{\begin{multline}}
\newcommand{\emult}{\end{multline}}
\newcommand{\eea}{\end{eqnarray}}
\newcommand{\ord}{{\cal O}}
\newcommand{\sumint}[1]

\begin{document}

\title{$O(N)$ symmetry-breaking quantum quench: 
Topological defects versus quasiparticles}  

\author{Michael Uhlmann$^{1}$, Ralf Sch\"utzhold$^{2}$, and 
Uwe R.~Fischer$^{3,4}$}  

\affiliation{
$^1$Department of Physics and Astronomy, University of British Columbia, 
Vancouver B.C., V6T 1Z1 Canada 
\\
$^2$Fachbereich Physik, Universit\"at Duisburg-Essen, D-47048
  Duisburg, Germany \\
$^3$Eberhard-Karls-Universit\"at T\"ubingen,
Institut f\"ur Theoretische Physik, 
D-72076 T\"ubingen, Germany \\
$^4$Seoul National University, Department of Physics and Astronomy, 
151-747 Seoul, Korea
}

\begin{abstract} 
We present an analytical derivation of the winding number 
counting topological defects created by an $O(N)$ symmetry-breaking 
quantum quench in $N$ spatial dimensions.  
Our approach is universal in the sense that we do not employ any 
approximations apart from the large-$N$ limit. 
The final result is nonperturbative in $N$, i.e., it cannot be 
obtained by 
an expansion in $1/N$, and we obtain far less 
topological defects than quasiparticle excitations, 
in sharp distinction to previous, 
low-dimensional investigations. 
\end{abstract} 


\maketitle

\section{Introduction}
%
In contrast to the vast amount of literature regarding {\em static} 
properties (e.g., universal scaling laws) of phase transitions -- 
both thermal and at zero temperature -- we are just starting to 
understand their {\em dynamical} features, especially the behaviour 
during a time-dependent sweep (quench) through the critical point. 
This topic has attracted increasing interest in recent years, see,  e.g., \cite{ZDZ,Damski,Dziarmaga,Uhlmann,DamskiZurek07,Lamacraft,Saito,Altland,Gritsev}
and, again, some universal properties became evident \cite{RS-LT}. 
For example, during a symmetry-breaking (second-order) dynamical phase 
transition, the diverging response time inevitably entails 
nonequilibrium processes and so the initial quantum (and thermal) 
fluctuations are amplified strongly, ultimately determining the final 
order parameter distribution. 
If the final phase permits topological defects 
(e.g., vortices in superfluids), they will generally 
be created in such a quench via (the quantum version of) the 
Kibble-Zurek mechanism. 
The latter occurs in many diverse physical settings, for instance,
in nonequilibrium phase transitions during the early universe 
\cite{Kibble,Boyanovsky,Stephens} or in condensed matter systems 
\cite{Zurek}.

Unfortunately, due to the inherent computational complexity of such 
scenarios, explicit calculations are difficult in general,  
and thus often rather uncontrolled assumptions and approximations 
(e.g., Gaussianity \cite{Liu,Lamacraft}) have been invoked.  
For example, the correlation function after the transition has been 
used to infer the number of created quasiparticle excitations 
(see, e.g., \cite{Polkovnikov}). 
The quasiparticle number is, then, supposed 
to directly yield an estimate for the topological 
defect densities generated by the quench.
For special cases such as the (exactly solvable) one-dimensional 
quantum Ising model (where the only excitations are topological 
defects, i.e., kinks \cite{ZDZ,Dziarmaga,Sen}), 
such an approach might give the correct answer -- but in general, 
this will not be the case, as we will argue below. 

In the following, we consider a rather general $O(N)$-symmetry 
breaking quantum quench and study the creation of topological 
defects (hedgehogs in the case considered) via calculating their winding 
number. 
In order to base our derivation on a well-defined expansion, 
we consider the large-$N$ limit.
Apart from the large-$N$ limit, no further approximations will 
be needed, i.e., our results will be quite universal. 
Moreover, similar to analogous large-$N$ approaches in condensed 
matter and field theory (assuming that there is no critical value
of $N$ where the system changes drastically), 
we expect our results to apply {\em qualitatively} also to 
finite $N$ (e.g., $N=3$), which are accessible to experimental 
tests.
Bose-Einstein condensates, in particular, permit the time-resolved
observation of the defect formation mechanism due to the 
comparatively long req-equilibration time scales of these dilute ultracold quantum  
gases \cite{Sadler,Weiler}.

\section{Effective action}
%
As a first step, we construct 
a general effective action 
for an $O(N)$-model in terms of the $N$-component field 
$\f{\phi}=(\phi_1,\dots,\phi_N)$, which determines the  
order parameter.
To this end, we start from the equation of motion with an 
arbitrary function $\f{f}$
\bea
\label{eom}
\f{\ddot\phi}=\f{f}(\f{\phi},\f{\dot\phi},\na^2\f{\phi},
\na^2\f{\dot\phi},\na^4\f{\phi},\na^4\f{\dot\phi},
\dots) 
\,.
\ea
In order to avoid run-away solutions and to facilitate a proper
quantum description, we have assumed the absence of time derivatives 
of third or higher order.
The initial state (before the transition) obeys the $O(N)$ 
symmetry: $\langle\hat\phi_a\rangle=0$ and 
$\langle\hat\phi_a(x)\hat\phi_b(x')\rangle\propto\delta_{ab}$,
etc. 
As stated, in all of our calculations, we employ the 
large-$N$ limit assuming $N\gg1$.
In this case, $O(N)$ invariant combinations such as 
$\f{\hat\phi}^2=\hat\phi_1^2+\dots+\hat\phi_N^2$ 
are sums of many independent quantities on an equal 
footing \cite{UV}.  
Considering commutators of such combinations, we obtain the 
well-known fact that their leading contribution (in the large-$N$ limit) 
behaves as a c-number whereas the (classical and quantum) fluctuations 
scale with $\sqrt{N}$ (cf.~the law of large numbers).  
Therefore, we may approximate 
\bea
\label{mean-field}
\f{\hat\phi}^2=\langle\f{\hat\phi}^2\rangle+\ord(\sqrt{N})
\,,\;
\langle\f{\hat\phi}^2\rangle=\ord(N)
\,,
\ea
arriving at a semi-classical (mean-field) expansion valid in the 
large-$N$ limit.
%
As a result, we may approximate the nonlinear terms in the 
equation of motion \eqref{eom}, for example  by 
$\f{\hat\phi}^3\approx\langle\f{\hat\phi}^2\rangle\f{\hat\phi}$,
arriving at a linearized description. 
This leads us to 
the most 
general linear and local $O(N)$ invariant 
effective action
containing up to first time derivatives
of the fields $\f{\phi}=(\phi_1,\dots,\phi_N)$ 
\bea
\label{action}
{\cal L}
&=&
\frac12\left(
\f{\dot\phi}\cdot F(-\na^2)\f{\dot\phi}
-\f{\phi}\cdot G(-\na^2)\f{\phi}
\right),
\ea
with arbitrary Fourier space functions $F(k^2)$ and $G(k^2)$.  

\section{Phase transition} 
\begin{figure}[b]
\psfrag{omega}{\large $\omega^2$}
\psfrag{k*}{\large $k_*$}
\psfrag{k}{\large $k$}
\centerline{\epsfig{file=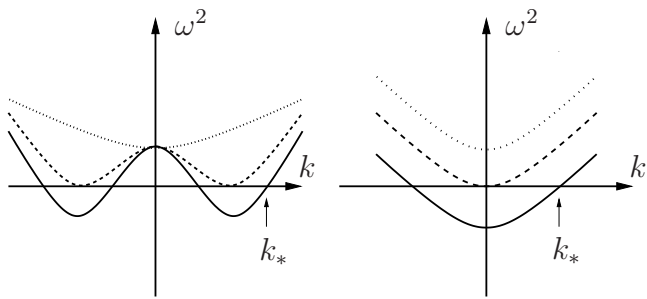,width=0.475\textwidth}}
\caption{\label{Fig1}
Two generic examples for the evolution of the dispersion relation 
\eqref{dispersion} during a symmetry-breaking phase transition, 
see also \cite{RS-LT}. 
Initially (dotted line), all $k$-values are stable, $\omega^2(k)>0$.
At the critical point (dashed line), the dispersion relation
touches the $k$-axis, and after the transition (solid line),
modes in a finite $k$-interval become unstable, $\omega^2(k)<0$.
The left panel corresponds to a case where $\omega^2(k=0)=m^2$ 
in Eq.\,\eqref{dispersion} remains positive while $c^2$ changes 
sign (see, e.g., \cite{supersolid}); whereas, in the right
panel, $\omega^2(k=0)=m^2$ becomes negative. 
In both cases, however, there is a dominant wave vector $k_*$
(for large $N$),
as indicated by the vertical arrows.}
\end{figure}

%
From Eq.~\eqref{action}, we derive a Klein-Gordon type
dispersion relation [to $\ord(k^2)$] 
for the linearized fluctuations, 
\bea
\label{dispersion}
\omega^2(k)=\frac{G(k^2)}{F(k^2)}=
m^2c^4 +c^2k^2+\ord(k^4)
\,.
\ea
Initially, all modes are stable, $\omega^2(k)\geq0$, since we
linearize around the initial [$O(N)$-symmetric] state.
After the $O(N)$-symmetry breaking transition, however, the state 
$\langle\f{\hat\phi}\rangle=0$ is no longer stable 
and the system ``wants'' to roll down to a state with 
$\langle\f{\hat\phi}\rangle\neq0$.
Typically (for second-order transitions, i.e., without meta-stability),
this implies that some of the modes become unstable, $\omega^2(k)<0$,
cf.~Fig.\,\ref{Fig1}. 
Since Eq.\,\eqref{action} is already a result of the large-$N$ limit, 
we assume that $\omega^2(k)$ is independent of $N\gg1$ 
(otherwise the group and phase velocities would either 
diverge or vanish 
{in the limit $N\to\infty$}). 
Furthermore, modes with sufficiently large $k$ should be stable
$\omega^2(k\uparrow\infty)>0$, so that the unstable interval in 
which $\omega^2(k)<0$ is assumed to be finite. 

\vspace*{1em}
\vspace*{-1em}

{So far, our results were independent of the number $D$ of 
spatial dimensions due to isotropy.
In the following, we set $N=D$ in order to facilitate the creation 
of topological defects in the form of hedgehogs (see below). 
So strictly speaking, we consider the simultaneous limit 
$N\to\infty$ and $D\to\infty$ and assume that these limits 
commute.
Since all relevant quantities such as 
$\langle\hat\phi_a(\bm r,t)\hat \phi_b(\bm r',t) \rangle$
depend on $|\bm r-\bm r'|$ only (isotropy), the large-$D$ limit
basically just affects the integration measure $d^Dk$, which 
strongly supports this assumption.
With a Fourier expansion of Eq.~\eqref{action}, we obtain the 
two-point function after the quench}
%
%
%
%
\begin{multline}
\langle\hat\phi_a(\bm r,t)
\hat \phi_b(\bm r',t) \rangle 
\\
=\frac{\delta_{ab}}{ (2\pi)^{N/2}}
\int dk\,k^{N-1}\,\frac{ J_\nu(k L) }{ (kL)^\nu }
\left[C_k^\pm e^{\pm2i\omega_kt}+D_k\right], \label{two-point}
\end{multline} 
with $L=|\bm r-\bm r'|$. 
The Bessel functions $J_\nu$ with index $\nu = N/2-1$ arise from 
the integration over all $\f{k}$-directions and the factors 
$C_k^\pm$ and $D_k$ depend on the initial state (for example the temperature) 
as well as quench dynamics, and are roughly independent of $N$. 
As expected, we obtain an exponential growth of the unstable modes, 
which have  
$\omega_k^2<0$, after the phase transition -- which then seeds 
the creation of topological defects. 
Of course, due to the growing modes, the linearization in 
Eq.~\eqref{action} will fail eventually -- but for $N\uparrow\infty$,
the time $t$ until which the linearization and thus 
Eq.~\eqref{two-point} applies does also grow. 
%
{Therefore, we may distinguish basically three phases following the 
quench:
First, we get a period of exponential growth of the modes where 
the linearized description in Eqs.~\eqref{action} till \eqref{two-point}
applies. 
Then, nonlinear effects set in and lead to a saturation of this growth 
and possibly an oscillation around the new energy minimum. 
Finally, the topological defects created by the quench start to 
``feel'' the attraction between hedgehogs and anti-hedgehogs 
leading to their approaching each other and eventual annihilation. 
This general picture has been qualitatively confirmed by numerical 
simulations \cite{Horiguchi} for $N=2$. 
The topological defects are seeded in the first phase 
(exponential growth) and slowly disappear in the final phase.
Therefore, we expect to obtain a good estimate for the maximum 
number of created defects at long length scales {and intermediate times}
from our linearized analysis. 
}

{In addition to the exponentially growing (in $t$) modes at finite 
$k$, the integral \eqref{two-point} does also yield a huge contribution 
from large $k$, because the phase space factor $k^{N-1}$ rapidly 
rises with $k$ for large $N$.
This gives rise to a strong $UV$ singularity of the two-point function 
$\propto |\bm r-\bm r'|^{-\ord(N)}$, see the discussion below of regularizing
this UV divergence.}

%

\section{Topological defects}
%
After the symmetry-breaking transition, the ground state is degenerate and 
can be specified by a nonvanishing expectation value 
$\langle\f{\hat\phi}\rangle\neq0$, which singles out a preferred direction 
given by the unit vector 
$\f{n}=\langle\f{\hat\phi}\rangle/|\langle\f{\hat\phi}\rangle|$.  
Thus the original $O(N)$ symmetry is broken down to $O(N-1)$, i.e., 
rotations around the $\f{n}$-axis, and the ground-state manifold 
corresponds to the surface $\mathcal S_{N-1}$ of a sphere in $N$ dimensions
$O(N)/O(N-1)\simeq\mathcal S_{N-1}$. 
Remembering the homotopy group $\pi_{N-1}(\mathcal S_{N-1})={\mathbb Z}$, 
we see that topological point defects in the form of hedgehogs exist in 
$N$ spatial dimensions \cite{Hopf,simple}. 
These defects correspond to nontrivial mappings from the 
ground-state manifold $\mathcal S_{N-1}$ onto the surface
$\mathcal S_{N-1}$ of a sphere in real space, characterized by a 
{winding number ${\mathfrak N}\in{\mathbb Z}$}, 
which reads \cite{Uhlmann,Abanov} 
\bea
\label{winding}
{\mathfrak N} =
\frac{\varepsilon_{abc...}
\varepsilon^{\alpha\beta\gamma...}}
{\Gamma(N)\parallel\!\mathcal S_{N-1}\!\parallel}
\oint dS_\alpha
n^a
(\partial_\beta n^b)
(\partial_\gamma n^c)...
\,,
\ea
where $\parallel\!\mathcal S_{N-1}\!\parallel = 2 \pi^{N/2}/\Gamma(N/2)$ 
is the surface area of the unit sphere in $N$ dimensions. 
Starting with the $O(N)$-symmetric state as the initial state,
we cannot simply insert 
$\f{n}=\langle\f{\hat\phi}\rangle/|\langle\f{\hat\phi}\rangle|$
since $\langle\f{\hat\phi}\rangle$ vanishes.
Therefore, we use a quantum operator $\f{\hat n}$ instead, 
which must be defined appropriately, and allows for a derivation
of the probability distribution of the 
quantum winding number $\hat{\mathfrak N}$ 
in a given volume from the above general expression. 
In particular, the expectation value of the winding number is of 
course zero,  $\langle\hat{\mathfrak N}\rangle=0$, but its variance
$\langle\hat{\mathfrak N}^2\rangle$ is in general not. 
%
{Setting directly $\f{\hat n}\propto\f{\hat\phi}$, we see that 
the variance $\langle\hat{\mathfrak N}^2\rangle$ is plagued with UV 
divergences similar to other quantities containing products of 
quantum fields at the same space-time point.
This is due to the fact that quantum fluctuations of $\f{\hat\phi}$ 
at arbitrary $k$-scales [cf.~the strong UV singularity mentioned 
after Eq.\,\eqref{two-point}] would in general contribute to 
$\langle\hat{\mathfrak N}^2\rangle$.}
In fact, even the $O(N)$-symmetric initial ground state can be 
viewed as a ``quantum soup'' of virtual hedgehog--anti-hedgehog 
pairs which are constantly popping in and out of existence.
Here, we are not interested in those virtual short-lived defects, 
but in long-lived hedgehogs, which are created by the quantum quench. 
Therefore, we 
{have to} insert a time-averaged unit vector 
defined via 
\bea
\label{time-average}
\f{\hat n}(\bm r)=\frac{1}{Z}\,\int dt\,g(t)
\f{\hat\phi}(t,\bm r)
\,,
\ea
with a smooth smearing function $g(t)$ and the normalization 
$Z=\langle[\int dt\,g(t)\f{\hat\phi}(t,\bm r)]^2\rangle^{1/2}
+\ord(\sqrt{N})$, 
where we have used the large-$N$ (mean-field) expansion.  
This time-average now suppresses all (rapidly) oscillating modes 
with $\omega_k^2>0$ and only leaves the growing modes $\omega_k^2<0$.
%
{After this UV-regularization, the integral in the two-point 
function \eqref{two-point} will be dominated by only a few modes 
in the vicinity of a certain wavenumber $k_*$:}
%
%
In view of the phasespace factor $k^{N-1}$ in \eqref{two-point}, 
the dominant contribution for large $N$ 
{\cite{remark}} 
will arise from the largest $k$ value for which $\omega_k^2 < 0$, 
i.e., close to the zero of $\omega_k^2$, cf.~Fig.\,\ref{Fig1}.
%
{Thus,} we can evaluate \eqref{two-point} in saddle point
approximation and obtain the correlator for the time-averaged 
direction vector 
{in \eqref{time-average}}
\bea
\label{Bessel}
\langle\hat n_a(\bm r)\hat n_b(\bm r')\rangle
 = 2^\nu \frac{\Gamma(\nu+ 1)}{N}
\frac{J_\nu(k_* L)}{(k_*L)^\nu} \delta_{ab}
= f(L) \delta_{ab} \,,
\ea
%
where higher-order terms $\propto N^{-3/2}$ stemming
from the normalization $Z$ 
{in \eqref{time-average}} 
have been omitted.
These higher-order terms vanish for $N\to\infty$
so that \eqref{Bessel} becomes indeed exact and, in view of
the asymptotic behavior of the Bessel functions \cite{Abramowitz},
a Gaussian correlator
\bea
f(L) = \frac{1}{N} \exp\left\{- \frac{k_*^2 L^2}{2N} \right\}
\label{Gauss}
\ea
follows.
Thus, the typical linear domain size (correlation length) 
is given by $L_{\rm corr}=\ord (\sqrt N/k_*)$.  
At extremely large distances $L=\ord(N/k_*)$ (where the first 
nontrivial zero of the Bessel function $J_\nu$ is located), there are 
oscillatory deviations, but in this regime, the correlator 
is already exponentially small.  
We emphasize that the 
Gaussian form of $f(L)$ stems from the large $N$ limit 
of the  exact expression in \eqref{Bessel}, 
and is not assumed {\em a priori}. 
Furthermore, as may already be observed in Eq.~(\ref{Bessel}), 
the emergence of a dominant wavevector $k_*$ implies the 
cancellation of all time-dependence, i.e.,
the time-dependence of the growing part of \eqref{two-point}
approximately separates such that
the time-averaged 
unit vector (\ref{time-average}) becomes independent of $g(t)$  
and thus stationary (in the regime under consideration). 
%
{So the emergence of a dominant scale in the correlator, $k_*$, {\em a posteriori} 
justifies the introduction of the UV regulator $g(t)$, which only 
affects the rapidly oscillating modes at larger $k$ but not the 
observables we are interested in.}


\section{Scaling laws}
%
Now we are in a position to derive the dependence of 
$\langle\hat{\mathfrak N}^2\rangle$ on $N$ and the enclosed volume.
Inserting Eq.~(\ref{time-average}) into the winding number variance 
$\langle\hat{\mathfrak N}^2\rangle$ from Eq.~(\ref{winding}), 
we obtain the expectation value of the product of $2N$ fields 
$\hat\phi_a$, which factorizes into $N$ two-point functions 
(\ref{Bessel}). 
Since these functions are completely regular, we may apply Gauss' law 
to the two surface integrals occurring in 
$\langle\hat{\mathfrak N}^2\rangle$, and get after some algebra
\cite{lang}
\begin{multline}
\label{algebra}
\left\langle\hat{\mathfrak N}^2\right\rangle
= \frac{N N!}{||\mathcal S_{N-1}||^2}
\int d^Nr\,d^Nr' \frac{1}{L^{N-1}} \frac{\partial}{\partial L}
\left(-\frac{\partial f}{\partial L}\right)^N 
.
\end{multline}
For a sphere of radius $R$, $V=\{\bm r\,:\,\bm r^2<R ^2\}$, 
we can evaluate this expression and finally obtain a single 
integral of the form
\bea
\left\langle \hat{\mathfrak N}^2\right\rangle
= 
\frac{N!}{\pi}\, R^N 
\int\limits_0^{\pi/2}\! d\theta  
\left(-\cos \theta \frac{\partial f}{\partial L}(2R \sin\theta)  
\right)^N 
\label{scaleint} 
.
\ea
Note that these 
{general expressions \eqref{algebra} and \eqref{scaleint}}
for $\langle\hat{\mathfrak N}^2\rangle$
are neither restricted to the Bessel functions \eqref{Bessel}, nor
to large $N$, but might be employed for any correlator of the form
$\langle \hat n_a(\bm r) \hat n_b(\bm r)\rangle = \delta_{ab} f(L)$
in any dimension $N \geq 2$.
%
{Let us discuss the scaling of 
$\langle\hat{\mathfrak N}^2\rangle$ with respect to $R$ and $N\gg1$, 
using the Gaussian correlator (\ref{Gauss}).
For radii far above the correlation length 
$R\gg L_{\rm corr}=\sqrt{N}/k_*$, the winding number variance 
\eqref{scaleint} behaves as} 
\bea
\left\langle\hat{\mathfrak N}^2\right\rangle
& = & 
\left( 
e^{-3/2}
\frac{k_* R}{\sqrt N} 
\left[1 + \ord(1/\sqrt{N})\right]
\right)^{N-1}
\label{scaleN} 
.
\ea
We observe that $\langle\hat{\mathfrak N}^2\rangle$ scales with 
the area $R^{N-1}$ of the hyper-surface enclosing the defects.
Apart from the prefactor $e^{-3/2}/\sqrt{N}$, this area scaling
is quite universial as it holds for spherical volumes in any
dimension $N \geq 2$ -- provided we assume short-range correlations 
-- and can already be inferred from Eq.\ \eqref{winding}:
If we calculate $\langle\hat{\mathfrak N}^2\rangle$ using 
\eqref{winding}, we obtain two hyper-surface integrals.
Due to isotropy, the first one yields $R^{N-1}$ while the second 
integral averages over the distance $|\bm r-\bm r'|$ between the 
two points on the surface.
Assuming short-range correlations only, this second integral 
becomes independent of $R$ (for large $R$) and gives  
$h(N) k_*^{N-1}$ with some function $h(N)$.
Note, however, that the assumption of short-range correlations
is crucial and nontrivial in this argument: 
For vortices in two dimensions, for example, we obtained 
logarithmic corrections to the ``area'' scaling,  
$\langle\hat{\mathfrak N}^2\rangle\propto R\ln R$ \cite{Uhlmann}, 
since the correlator fell off quite slowly at large $L$. 

{If the radius $R$ shrinks and approaches the correlation 
length $R\sim L_{\rm corr}=\ord (\sqrt N/k_*)$, the winding number 
variance decreases rapidly (for $N\gg1$). 
A sphere with $R=\ord (\sqrt N/k_*)$ would then contain around one 
defect (or anti-defect) on average, 
$\langle\hat{\mathfrak N}^2\rangle=\ord(1)$, which determines 
the {\em total} defect density. 
For even smaller radii, far below the correlation length 
$R\ll L_{\rm corr}=\ord (\sqrt N/k_*)$, the above formulae 
would yield a scaling $\langle\hat{\mathfrak N}^2\rangle\sim R^{2N}$,
i.e., an exponential suppression (for large $N$).
However, the precise functional form \eqref{scaleint} should not be 
trusted upon in this regime since we have neglected 
$\ord(1/\sqrt{N})$-corrections in our derivation, which is  
problematic if the final result is exponentially small.}
From  a more physical point of view, the mean-field approximation 
\eqref{mean-field} breaks down near the core of a defect
(where $\f{n}$ becomes ill-defined), which renders 
Eq.~\eqref{scaleint} questionable for too small volumina. 
{For small $R$, one would expect a volume-type scaling
$\langle\hat{\mathfrak N}^2\rangle \sim R^{N}$, i.e.,
the typical behaviour for uncorrelated defects,
which should be the case if there is one hedgehog at most.
The exponential suppression
$\langle\hat{\mathfrak N}^2\rangle \sim \exp\{-\ord(N)\}$
for large $N$ and small $R \ll L_{\rm corr}$ should still
be correct, as this just reflects the diminishing probablity
of reversing field orientation in {\em all} directions when
increasing $N$.
}

%
%
%
%

The area scaling \eqref{scaleN} of the {\em net} defect number
$\langle \hat{\mathfrak N}^2\rangle$ can be interpreted as the
occurrence of a confined phase of bound defect-antidefect pairs.
%
Only pairs where one of the partners is contained within while the
other is outside the integration volume would yield net winding
number, whereas those pairs entirely inside or outside do not
contribute.
Hence a scaling with surface area instead of volume is natural 
for short-ranged correlations.
On the other hand, there could, in principle, also exist a
de-confined phase of quasi-free hedgehogs similar to the quark-gluon
plasma of quantum chromodynamics.
Such a de-confined phase might occur if defect density and
temperature are sufficiently high and any bound pairs are
broken up again by thermal quasi-particles.
In that case, defects and antidefects would be randomly distributed
and a volume scaling of the winding number variance
$\langle \hat{\mathfrak N}^2 \rangle$ follows.
Since we did not make any assumption in our derivation apart from
the large $N$ limit (where the results become exact), we can clearly
distinguish between the two phases (confined or de-confined).
%

\section{Statistics}
%
In a similar manner, 
we can calculate the higher moments of the winding number.  
Again, by exploiting the fact that we have short-range correlations,
the large-$R$ limit of the next nontrivial moment can be inferred 
from pure combinatorics in the analysis of the four integrals
occurring in   
\bea
\langle\hat{\mathfrak N}^4\rangle
=
3\langle\hat{\mathfrak N}^2\rangle^2
+\ord(R^{N-1})
=
{\ord(R^{2N-2})}
\,.
\ea
Analogously, the leading terms of 
$\langle\hat{\mathfrak N}^{2n}\rangle$
are given by $(2n-1)!!\langle\hat{\mathfrak N}^{2}\rangle^n$
with $(2n-1)!!=(2n-1)(2n-3)\dots5\cdot3$. 
For large $R$, the winding number $\hat{\mathfrak N}\in\mathbb Z$
can be approximated by a continuous variable 
$\hat{\mathfrak N}\in\mathbb R$ and thus its full statistics 
is given by the inverse Mellin transform of 
$(2n-1)!!=2^n\Gamma(n+1/2)/\sqrt{\pi}$, which yields the 
Gaussian probability distribution 
$p({\mathfrak N})\propto\exp\{-\gamma^2{\mathfrak N}^2\}$, 
with $1/\gamma^2=2\langle\hat{\mathfrak N}^2\rangle$. 
We note that, like in Eq.~\eqref{Gauss}, the
Gaussianity is not assumed but derived from 
first principles in a given limit -- 
for small $R$ (small ${\mathfrak N}$), for example, there will be deviations
from a Gaussian distribution.  

\section{Conclusions}
%
Based on a very general $O(N)$-invariant effective action, we presented 
an analytical derivation of the winding number counting the defects 
created by a symmetry-breaking quantum quench in the large-$N$ limit. 
Consistent with previous calculations \cite{Rajantie}, 
our result \eqref{scaleN} is nonperturbative, i.e., 
it does not admit a Taylor expansion in $1/N$.
As another result, we find that the typical distance between defects 
scales with the correlation length $\ord(\sqrt{N}/k_*)$. 
By contrast, the typical distance between quasiparticle excitations 
(e.g., Goldstone modes) does not increase with $N$.
This can be understood by recalling that the total energy of the system 
(which scales with $N$) in a given volume has to be distributed among  
all the quasiparticle excitations, whose typical 
energy is determined by the dispersion relation $\omega^2(k)$ and thus 
independent of $N$.
Therefore, we conclude that 
the quasiparticle spectrum alone does not yield any direct 
information about the generation of topological defects in general.
%
This situation is quite different in the one-dimensional 
quantum Ising model, where topological defects (kinks) are the 
only quasiparticle excitations \cite{ZDZ,Dziarmaga}, 
which frequently led to the assumption in the literature that 
this is generic. 
We demonstrated here that identifying quasiparticle excitation and defect 
numbers created by a quantum quench can be quite misleading. 

The crucial difference between quasiparticles (whose number can be 
derived via a perturbative expansion in $1/N$) and  topological defects 
(which are nonperturbative) can be illustrated by the 
following intuitive picture:
Considering a discrete regular lattice with a unit direction vector 
$\f{n}_i$ at each lattice site $i$, a quasiparticle excitation occurs
if $\f{n}_i\neq\f{n}_j$ for two neighbours $i,j$.
A topological defect at the site $i$, one the other hand, means 
that the unit vectors $\f{n}_j$ of {\em all} neighbouring sites 
either point away or towards the site $i$.
For large $N$, this is obviously a much stronger condition. 

%
%
%

It is also worth noting that the derived area scaling 
$\langle\hat{\mathfrak N}^2\rangle\propto R^{N-1}$ 
is inconsistent with the random defect gas model 
(where defects and anti-defects are distributed randomly 
in the sample volume, corresponding to a de-confined phase) 
since this model would predict a volume law, i.e., $R^N$-scaling.
We remark in this connection that the area scaling 
{\cite{Eisert}} 
(corresponding to a confined phase) 
we obtain can be interpreted by a random $\f{n}$-field model on 
the hyper-surface with the correlator Eq.~(\ref{Bessel}), representing 
a generalization of the random phase walk model for 
$N=2$ (cf.\,the result of \cite{Uhlmann} in which reasonable
agreement with the  experiment reported in \cite{Sadler} was obtained). 
 
Finally, we would like to stress that our result is quite universal, 
i.e., it is valid for very general dispersion relations 
of the $O(N)$ 
model (cf.~Fig.\,\ref{Fig1}) and just relies on the large-$N$ limit without
any further approximations.
Moreover, as indicated below Eq.\,\eqref{scaleN}, we expect that the 
general picture does still apply qualitatively for smaller, 
and thus experimentally accessible values of $N$, for example $N=3$.
In particular, this should be true for fast quenches, where we have 
a well-defined period of exponential growth of the unstable linear 
modes, while nonlinear effects (saturation of this growth, 
oscillations, and finally defect annihilation, see \cite{Horiguchi}) 
occur much later. 
In this case, one may find (instead of $1/N$) another small parameter 
(e.g., the diluteness of the gas) in order to motivate the underlying 
effective action in analogy to Eq.\,\eqref{action}.
For $N\not\gg1$, universality will be partially lost and the 
dependence on the dispersion relation, for example, will be stronger.
For instance, it might then be necessary to introduce a time-dependent 
critical $k_*=k_*(t)$, which is 
not close to the zero of $\omega^2(k)$, but near the actual minimum of 
$\omega^2(k)$. 

{\acknowledgments}
M.\,U. acknowledges support by the 
Alexander von Humboldt Foundation and NSERC of Canada, 
R.\,S. by the DFG (SCHU~1557/1-3, SFB-TR12), and 
U.\,R.\,F. by the DFG (FI 690/3-1) and the 
Research Settlement Fund of Seoul National University. %
\vspace*{-0.2em}


\begin{thebibliography}{499}
\bibitem{UV} 
Of course, for quantum fields, this expression 
must be UV-regularized to render it well-defined, 
cf.~Eq.\,\eqref{time-average}.
\bibitem{simple} 
Due to $\pi_{M-1}(\mathcal S_{N-1})=\emptyset$ for $M < N$, 
there are no other topological defects (such as vortex lines, 
belonging to the fundamental group $\pi_1$) 
apart from the hedgehogs.  
%
The relation $\pi_{M-1}(\mathcal S_{N-1})=\emptyset$ reflects the 
fact that mappings of a given $\mathcal S_{M-1}$ to the surface 
of a sphere of higher dimension $\mathcal S_{N-1}$ are always 
contractible to a point and are thus topologically trivial \cite{Hopf}, 
a simple example being a rubber band ($\mathcal S_{1}$) on a bowling ball 
($\mathcal S_{2}$), which may slip off easily.
%
{In contrast, the first homotopy group of the torus $\mathcal T$ is 
nontrivial, $\pi_{1}(\mathcal T)={\mathbb Z} \times {\mathbb Z}$ 
[i.e., is characterized by two integer winding numbers], and hence it is 
possible to wind the rubber band ($\mathcal S_{1}$) around the torus 
such that it cannot slip off.}
\end{thebibliography}
\end{document}